# SAMPLE – stratospheric altitude microbiology probe for life existence – a method of collection of stratospheric samples using balloon-borne payload system

## Abstract

The Earth possesses many environmental extremes that mimic conditions on extraterrestrial worlds. Earth's stratospheric conditions at 30–40 km altitude are very similar to the surface of Mars: with same pressure, average temperature, and even same levels of solar UV and proton radiation, and Galactic cosmic rays. Microbial (bacteria and fungi) habitation in troposphere is known and well documented, however, very little is known about the true upper limit, or extent, of the Earth's biosphere. Stratosphere provides a good opportunity to study the existence, or the survival, of life in these conditions. Despite the importance of this topic to astrobiology, stratospheric microbial diversity/survival remains largely unexplored, probably due to significant difficulties in the access and in ensuring the absence of contamination. To conduct a detailed study into this, we have designed the balloon-borne payload system SAMPLE (Stratospheric Altitude Microbiology Probe for Life Existence) to collect dust samples from stratosphere and bring them in to a suitable environment, where further study will be conducted in establishing the possibility of microbial life in the upper atmosphere. The entire design was created in-house and is entirely novel, with modifications made to the materials chosen for weight reduction and stress reduction of all kinds. The main payload consists of three pre-sterilized sampling trays, and a controller which will determine the altitude of the payload system to actively monitor the opening and closing of the collection trays. For additional contamination control, one of sampling trays will fly but not open, and one will remain closed on the ground in the cleanroom. Other on-board devices include environmental sensors, GPS tracking devices, cameras, and an FTU (Flight Termination Unit) to terminate the flight after the payload has reached the desired altitude and on completion of the sample collection. A parachute attached to the payload ensures safe descent of the payload on its way back to the surface. On retrieving the payload, the sampling trays (including controls) will be sent to a suitable laboratory where the samples will be examined for the nature of collected material.

**Keywords:** high-altitude balloons, stratosphere, astrobiology, biota

Margarita Safonova,[1] Bharat Chandra P,[2] Binukumar G Nair,[3] Akshay Datey,[4] Dipshikha Chakravortty,[5] Ajin Prakash,[6] Mahesh Babu,[7] Shubham Ghatul,[7] Shubhangi Jain,[7] Rekhesh Mohan,[7] and Jayant Murthy[7]

[1]M P Birla Institute of Fundamental Research, India
[2]Raman Research Institute (RRI), India
[3]Tata Advanced Systems Ltd., India
[4]Center for Biosystems Science and Engineering, Indian Institute of Science (IISc), India
[5]Dept. of Microbiology, Indian Institute of Science (IISc), India
[6]Digantara Aerospace Company, India
[7]Indian Institute of Astrophysics (IIA), India

**Correspondence:** Margarita Safonova, M P Birla Institute of Fundamental Research, Bangalore 560001, India

## Introduction

Earth possesses many environmental extremes mimicking conditions on other worlds. To understand the origin of life and to successfully look for it in the Universe requires comparative studies of such Earth-based natural environments – terrestrial analogues of extraterrestrial environments.[1] Earth is currently the only known to us habitable planet, but there is a good probability that life could have existed on Mars in the past when it had liquid water on the surface.[2,3] The discovery of a desert varnish on Mars,[4] believed to be the product of the specific bacteria on Earth,[5] has gotten us even further down the road of whether life existed or even exists now on Mars with its super extreme conditions for habitation. Short of going to Mars, the closest habitat conditions on Earth are in the upper layers of atmosphere – stratosphere – at altitudes of 20–50 km, see e.g.[6] Though there are many sites on the surface closely resembling Martian environments,[7] stratospheric physical conditions, UV radiation, humidity, temperature and nutrient deprivation, are most similar to the Martian surface (**Table 1**). We know that life on Earth is just everywhere we look, surviving and thriving in every possible niche, with or without sunlight, water, or oxygen. Some organisms make their own water, take energy from all possible resources, can hibernate for millions of years. Life exists and proliferates deep in the Earth – in the deep biosphere that extends to more than 5 km in the crust and up to 2.5 km beneath the ocean floor.[8]

**Table 1** Earth's physical stratospheric conditions are very similar to the conditions on Martian surface

| | Martian surface | Earth at 30 km altitude |
|---|---|---|
| Pressure | 6 mbar | 5–10 mbar |
| Average temperature | −55◦ | −50◦ |
| Magnetic field strength | ≈ 400 nT | ≈ 50 μT |
| Galactic cosmic rays | > 0.01/cm3/s | > 0.01/cm3/s |
| Solar proton radiation for energies up to 1 Mev | ≈ 1/cm2/s | > 1/cm2/s |
| UV intensity | ≈ 0.06 W/cm2 | ≈ 0.07 W/cm2 |
| Total solar intensity | ≈ 0.6 W/cm2 | ≈ 1.3 W/cm2 |

But is there life high up in the atmosphere? Some questions still remain unanswered:

a) Is there an upper limit to the biosphere?

b) What is the viability of life upwards?

c) What limits (or facilitates) upward spread?

d) Can life grow and reproduce in atmosphere, or is it just surviving?

Microbial (bacteria and fungi) habitation in troposphere (0–10 km) is known and well documented.[9] It is also known that bacteria continue to grow at a height of ~3 km (in supercooled ă 0″C cloud droplets).[10] But though samples of viable airborne microbes were previously brought from stratosphere,[11] some reports are not without controversy. Early reports from last century's 30s and even 70s cannot be verified on the subject of absence of contamination; spores brought from the altitude of 80 km by a rocket flight to the mesosphere[12] could have been from the rocket itself, or even the landing site in the Kazakhstan desert. Narlikar et al.,[13] reported the results of the balloon flight collection of air samples from 30–39 and 40–41 km altitudes as detection of extraterrestrial (ET) microorganisms raining constantly on Earth from space (embedded in comets/meteoroids). They reported that clumps of cocci-shaped sub-micron sized particles (as clumps of viable cells) fall on Earth at a rate of ~3 ton/day, as well as detection of a few culturable species.

Shivaji et al.,[14] reported detection of 12 bacterial strains at 27–41 km with 3 novel species of genera *Bacillus* and *Janibacteria*. Since one of the strains, *Bacillus Aryabhattai*, have not been previously detected on Earth, they assumed it to be ET. Subsequently, Ray et al.,[15] reported finding the *Bacillus Aryabhattai* strain in an isolate from East Kolkata wetlands. However, they also concluded that because this microbe has UV, salinity, Cr3` resistance, and ă 4″C preference: suited to growth at colder temperatures (no growth at ą 50″), it must be of extraterrestrial origin. At least, this controversy was resolved recently in the report where this bacillus was reclassified to be similar to a known terrestrial species.[16]

Wainwright et al.,[17] described several high-altitude balloon flights, where they claim to have discovered several 'weird' biological entities, including a titanium-shell bacterium and a diatom frustule, which presumably coming down to Earth embedded in cometary micrometeorites.[18] They, however, give little details of the payload and sampling chambers structure, and of the environmental conditions under which the payload was assembled and later disassembled to transport to the lab. At least for the diatoms, reports are available showing their widespread in the upper atmosphere.[19–21] In addition, only a broken part of a diatom frustule was found in July 2013 flight, and these are ubiquitous in any dusty environment — a diatomaceous earth.[22] In fact, that particular balloon launch was from a place near Ellesmere Port, Cheshire, UK, and the company Cheshire Pump Co. Ltd., located in Ellesmere Port, Cheshire, UK, is one of leading UK's supplier of diatomaceous earth pumps.

The major premise on which the above mentioned reports base their findings to be ET was that they could not conceive of any mechanism by which the viable terrestrial organisms can reach stratospheric altitudes, citing the tropopause as a natural barrier. However, it is well known that the height of the tropopause is not constant, either in space or in time, and there exist tropopause breaks where exchanges between the troposphere and stratosphere take place.[23] The tropopause itself has been continuously rising at 160 feet per decade since 1980s, and at nearly 180 feet after 2000, due to the anthropogenic climate change.[24] Normal cirrus and cumulonimbus clouds also extend much above the tropopause. In addition, a convective overshooting over tropical land is the key contributor to troposphere- to-stratosphere transport, injecting trace gases, ice water droplets and tropospheric air into the lower stratosphere.[25] Recent works suggested that upward vertical winds at altitudes of 90–250 km can transport biological particles into mesosphere and lower thermosphere,[26,27] which can explain the previous detections of bacterial and fungal spores at 77 km.[12]

On the other note, why would a space organism resemble, or look totally identical morphologically, to a species that evolved and living in a completely different environment (i.e. Earth)? Why should we assume that ET life would be identical to ours? There are more than 500 different amino acids in nature (for example, a Murchison meteorite contained 90 amino acids, out of which only 19 were known on Earth,[28] while terrestrial organisms use only 20 of them.

In addition, balloon flights described in Shivaji et al.,[14] were launched from the Indian TIFR Balloon Facility (TIFRBF) in Hyderabad. These are zero-pressure balloons capable to carry ~1 ton of payload, reaching ~50 km and floating for hours. To achieve that, the payload releases the ballast, which is about ~10% of the payload mass, to boost higher rise. The ballast module is located under the payload (Figure 1, *Left*), and the ballast is usually released gradually per packets of 1 kg.[29,30] Subsequently, the payload sucks in the air through the micropore filters for samples collection. Ballast, used in TIFRBF balloons, consists of tiny iron balls, which are obviously not sterilized prior to launch as it is impossible to do so (Figure 1, Right).

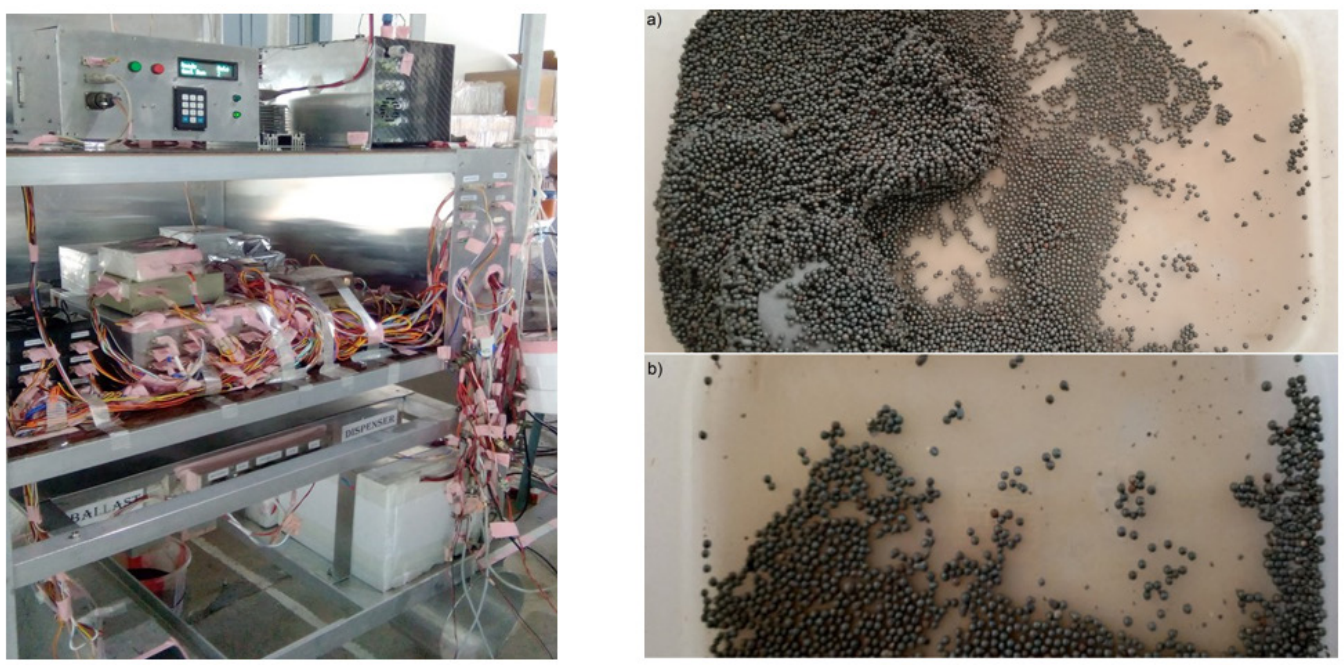

**Figure 1** (*Left*:) TIFRBF atmospheric experiment payload with the ballast dispenser module; the bucket filled with the ballast can be seen under the payload. (*Right*:) a) An actual photograph of the ballast used in TIFR zero-pressure balloons. Sizes of the balls range from 300 micron; b) A zoom of the top image. The photographs are from the flown experiment described in.[32]

We have established a high-altitude balloon program at the Indian Institute of Astrophysics (IIA), Bangalore, India in 2013,[31] and had many successful flights since then.[32] Due to regulations on flying balloons, we are constrained to fly only light-weight balloons with payloads not exceeding 5–6 kg. Our current goal is to design, construct and fly a Stratospheric Altitude Microbiology Probe for Life Existence (SAMPLE) payload on a high-altitude balloon to collect dust samples at altitude of ~30 km.

Along the design and development of a payload, we are establishing physical and biological procedures (including ground and positive control) for investigation of collected samples. We plan to use the scanning electron microscopy (SEM), and DNA metagenomic sequencing of samples suspected to be biological.

## Dust in Earth's stratosphere

Stratospheric dust originates from both space and from the ground. Most of the influx of extraterrestrial matter that falls onto the Earth is dominated by micrometeoroids with diameters in the range 50 to 500 micron, of average density 2.0 g/cm3 (with porosity about 40%). Though fine interplanetary dust (IDP) dominates this flux, still about 25 million meteoroids, micrometeoroids and other space debris particles enter atmosphere each day: remains of comets, asteroids, and even impact ejecta from the Moon, Mars, Venus, Mercury. The total flux of material arriving at Earth could be 30-40 tonnes per day.[33,34] In circumstellar dust, astronomers have found molecular signatures of CO, silicon carbide, amorphous silicate, polycyclic aromatic hydrocarbons, water ice, and polyformaldehyde, among others (in the diffuse interstellar medium, there is evidence for silicate and carbon grains).[35] and references therein. In fact, comets and asteroids are a rich source of organics delivery to Earth; the late heavy bombardment (LHB) ~4 Gigayears ago could actually have provided exactly the needed large chemical and thermal disequilibrium conditions for the biogenesis to (re)start. Meteorites from both comets and asteroids were found to contain amino acids, sugars, purines,[36] and, recently, a first protein was discovered in a chondritic meteorite found in the Sahara desert.[37] Organic material in chondritic meteorites, and to a lesser extent comets,[38] might have brought organics to the Earth at the rate of $10^{14}$ kg/yr, at a cumulative total several magnitudes greater than what might existed before.[39] Every year, hundreds of millions of tonnes of terrestrial dust are transported between continents through the air. This dust consists of soil particles, anthropogenic traces, all kinds of aerosols and, since microorganisms are virtually everywhere, this dust contains microorganisms – aerial microbiota, with concentrations reaching ~ $10^6$ $m^{-3}$ over land.[40] This microbiota plays an important role in the global climate system and atmospheric ecological dynamics.[41]

## Sources of airborne biomass

It is generally assumed that sources of microorganisms in stratosphere are arid top soil particles and marine sea spray in which microorganisms can survive being airborne. However, there are many ways by which terrestrial microorganisms reach higher altitudes in the atmosphere: injected into the stratosphere by dust storms, volcanic activity (fine particles from volcanic eruptions can reside in the stratosphere for up to 3 years), severe weather, and human activity – soil cultivation and forest burning. No surprise that most prevalent reported species was *Bacillus sub.* as these are ubiquitous in soil and marine waters. *Pseudomonas stru* is also present in virtually all environments. Large size of dust congregates can protect microbes from the UV, but will limit the residence time – the most relevant factor for stratospheric biomass estimates. The residence time of small (ă 10 $\mu$m) cells in stratosphere can range from months to years due to laminar flow of stratospheric winds.[42,43]

Increased human activity became the major source of bioaerosols in the stratosphere in the recent decades: with airplanes, rockets, weather and scientific balloons. In 2024, there was ~36.4 million commercial air flights globally (forecasted to reach 40 million in 2025), and 263 space launches (on average 120 space rocket launches a year since the 1960s)[1]. All this introduces countless scores of microbial life into the high atmosphere, and possibly even into the orbit, which was, for example, hinted at by the discovery of a plankton on the surface of the illuminator of the International Space Station (ISS).[44] Plus, our biosphere is already extended till at least the low-Earth orbit and, most probably, to the Moon.[45] In the recent years, the dramatically increased space debris adds to the introduction of biota to the stratosphere from the low-Earth orbit. There are estimated 128 million objects orbital space debris of particles in 1-mm to 1-cm size range.[46] Sources of this space debris are dead spacecraft, lost equipment, remains of weapons testing, etc., even human waste from space shuttles[47] and space stations[2]. Atmospheric drag causes the vast majority of the large debris to decay from orbit within a decade, creating small particles eventually descending down to the surface, which could be the source of the so-called 'ghost particles' detected by Wainwright et al.,[17] (Figure 2, *Middle*). In addition, the recent anthropogenic climate change increased the aerosol exchange in the tropopause, resulting in a greater number of microorganisms crossing into stratosphere.[48,49]

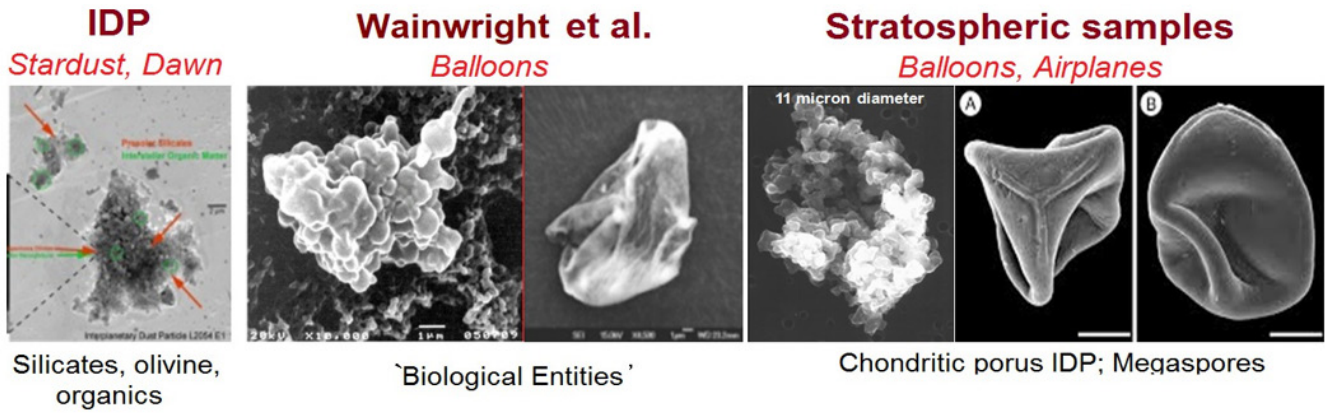

**Figure 2** Previous stratospheric isolates along with the 'iconic' IDP collected from space (*Left*). *Middle*: 'Biological entities' image from[17]. *Right*: Image of the chondritic porous IDP is from NASA Cosmic Dust collection (http://curator.isc.nasa.dust); and images of megaspores (A and B) lifted from the ground are from.[50]

In Figure 2, we show for comparison the dust particles collected from space (*Left*) and from stratosphere (*Right*), correspondingly. It may be noted that the so-called 'biological entities', reported by Wainwright et al.,[17] resemble morphologically both the IDPs from space and the organic particles lifted from the ground, thereby making their conclusions about the bio-ET origin of the isolates premature.

## Methods for sample collection

Many methods are employed to collect atmospheric samples of dust/biota. Cosmic dust has been successfully collected on the ground in the collecting pools, but the most common type of space dust found in these collectors are the particles containing iron. And the old method of melting snow from the rooftops in urban areas and draining the water past magnets[51] is still used successfully today by both amateur astronomers and research scientists.[52] and the Project Stardust (http://www.treasuresfromspace.com/). IDPs with sizes upto 1 mm have been recovered from melting the Antarctic ice, and the method of melting the fresh Antarctic snow rather than ice, to avoid micrometeorites alteration, has been described in Duprat et al.,[53] However, these kind of particles are usually already mixed with dust from the ground or the troposphere, therefore it is better to get higher into the atmosphere. The only way to go in order to prove or, at least, test the notion of lithopanspermia is to establish the regular collection and testing of the samples from the stratosphere. It is the first receiving region – it is the region where the meteoroids break up, the material spreads out and is not yet burnt, and are at lower speeds that in the LEO. Also, by the time the material rains down on the surface, it is already reprocessed and contaminated (as, for example, one of the most probable causes of finding biomaterial in the Polonnaruwa meteorite[54] was contamination with ground freshwater.[55]

The following methods were used to collect cosmic dust and estimate the rate at which cosmic dust rains on our planet.

A. **Space probes** Examples are NASA's Stardust spacecraft returning dust samples of the comet 81P/Wild-2 and, in 2014, the recovery of particles of interstellar dust from the Discovery program's Stardust mission.[56] Analysis of data from dust detectors aboard the Ulysses and Galileo spacecraft have revealed that there is a stream of inter- stellar dust flowing through our Solar System. These grains, of unknown mineralogy, are generally less than 1 $\mu$m and are very difficult to collect in the Earth's stratosphere by current techniques using high-altitude aircraft.

[1] Source: Global air traffic–number of flights 2004-2025, Statista Inc., ttps://www.statista.com/statistics/564769/airline-industry-number-of-flights/

[2] On March 12, 2021 *Spaceflight Now* reported release of 2.9-ton garbage load from ISS.[48] The garbage carrier is expected to remain in orbit for 2–4 years before burning out during atmospheric re-entry, which is the usual practice for garbage ejected by the ISS.

**B. Space stations**. Low-density silica aerogel was used since 1990s on the Mir Space Station,[57] and ISS later,[58] to capture the IDP grains. The grains are later removed from the aerogel, thin-sectioned using a diamond blade of an ultramicrotome, and imaged in a transmission electron microscope.

**C. LEO satellites/shuttles** Even in the early days of space shuttle flight there were measurements of the dust flux, e.g. the microabrasion foil experiment (MFE) on the STS-3 Space Shuttle mission.[59] Several active dust sensors were flown in the near Earth environment, such as e.g the Debris In-Orbit Evaluator (DEBIE)[60] and the Geostationary Orbit Impact Detector (GORID)[61] (for a good review, see Wozni- akiewicz et al.,[62]

**D. Meteorological rockets** Imshenetsky et al.,[12] described using meteorological rockets fitted with specially designed analyzers to collect samples at altitudes of 48 to 77 km.

**E. High-altitude balloons** For example, the DUSTER (Dust in the Upper Stratosphere Tracking Experiment and Retrieval) payload, flown on a zero pressure balloons.[63]

**F. Airplanes** Examples are NASA Cosmic Dust Program, employed with inertial-impact collectors mounted underneath the wings of high-flying (17 and 19 km) U2 and WB- 57F aircraft.[64] Smith et al.,[65] describes several flights with bioaerosol collectors till altitudes of 12.1 km, with the conclusion that bacteria in low stratosphere is not different from the tropospheric populations. This method, however, is limited by the upper limit of ~20 km, above which airplanes cannot fly.

### Collecting media

Till date, different collecting media and methods have been used in both space and in the atmosphere:

1) **Aerogel**. It is a silicon-based solid with a porous, sponge-like structure, in which more than 99% of the volume is open space. Aerogel is 1,000 times less dense than glass. This material has uniquely low thermal conductivity, refractive index, and sound speed, in addition to its exceptional ability to capture hyper-velocity dust. It was used in space collections by Stardust spacecraft.[56]

2) **Xerogel**. Similar to aerogel but with different preparation technique. Xerogel is denser, it was used in the 1999 collection of Leonids' dust by NASA stratospheric balloon flight.[66]

3) **Silicon grease/oil** (polydimethylsiloxane, $n(CH_3)_2SiO$). The first stratospheric IDPs were collected in 1970s using a balloon-borne collector called the Vacuum monster.[67] Particles were sucked in and collected by inertial impaction on the silicon- oil-coated glass rods. Since 1981, NASA has routinely used Si-oil-coated collectors deployed on high-altitude aircraft by inertial impaction.[68] In all such cases, particles were picked up selectively using a micro-manipulator, micro-needles and brush hair tools and, after washing the silicone oil with hexane, are photographed by SEM. This method has problems due to the Si contamination by the highly viscous silicon oil (particles may remain coated with it) and loss of organic material, if present, due to organic solvents, see e.g. Bradley et al.,[69]

4) **Polymer filter**. Volumes of air passed through few micron (0.2 $\mu$ or 0.45 $\mu$) pore size polymer filters. This requires very powerful differential pumping system with unidirectional valves to control the flow of pumped air. Since the air pressure decreases with the increase of altitude, the pumping system must be capable of producing high compression ratio (~1000) and low ultimate pressure. Such systems are achieved by employing turbo molecular pumps in differential pumping mode with multiple collecting chambers.[63]

5) **Polyurethane foam (PUF)**. Being a dry medium, it is more advantageous in comparison with the silicon oil method. The foam consists of a 3-D network of 40$\mu$-thick strands and thin membranes forming ~400$\mu$-wide cells. However, locating and extracting particles is difficult due to the complex structure of the foam.[70]

### Collecting media chosen for SAMPLE

Aerogel and xerogel are, of course, the best choices for the collection, however they are quite expensive. We have decided to use agar as a collecting medium. However, since the payload will be going through temperatures range of.20″ C to —60″ C and back, the agar gel coating may not survive this temperature cycling. The difference in the freezing points of agar and water may lead to disruption of agar suspension upon freezing and the water would ooze out of the system making the agar gel system collapsible.

The following experiment was set up to determine the optimum agar concentration required to maintain its architecture even after extreme freezing conditions. Different agar concentrations (2%, 5% and 10%) were prepared, poured on the tray, and kept at —80″ C cold chamber of the IISc Dept. of Microbiology for 24 hours (Figure 3(a)). Next, the tray was kept at room temperature allowing it to thaw. The amount of water lost after thawing was measured and inversely correlated to the extent of maintained agar architecture. We noticed that 2% and 5% agar preparations were unable to maintain their architecture, whereas 10% agar was found to be most stable after the freeze-thaw cycle (Figure 3(b)). It is interesting that if agar admixture is based not on pure water but on water with added glycerine, at glycerine concentration of 40%, even after freezing to —70″ C, there is no crystal ice formation after thawing.[71]

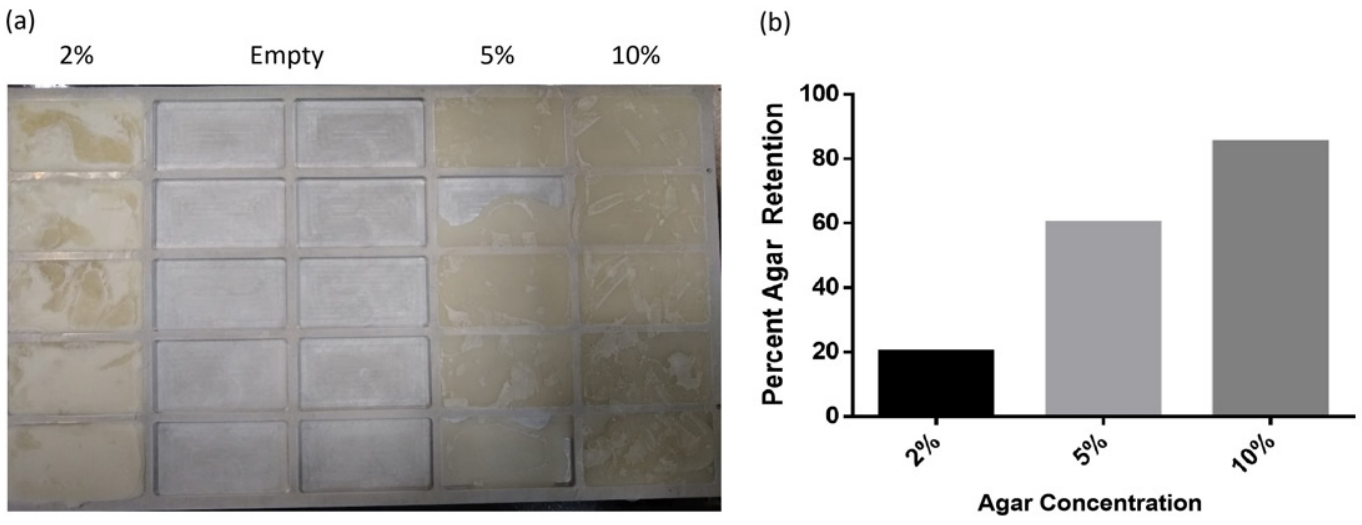

**Figure 3** Experiment with agar concentration. a) A tray with different agar concentrations coatings. b) Percentage of agar retention in a sample after thawing.

## Design overview

While designing the mechanical and electronic components, we need to consider several constraints imposed on the instruments. First, we are limited in weight of the overall payload at 6 kg – the limitation imposed by the Airport Authority of India. Secondly, since the temperatures in the stratosphere go down as low as —80″ C, all equipment must be able to withstand these temperature variations. Our initial design consisted of a box (Figure 4, Left) with 2 sterilizable rotating collecting chambers: one to open, and one control to keep closed, with detachable collecting plates coated with PUF. It included 2 rotating doors, 2 servo motors (D50024MG, torque 5.00 kg.cm (4.8 V)). However, we have ran into problems with the servo motors as they could not work at low temperatures, and if we were to include the heaters, the overall design became too heavy. Therefore, we had to come up with an innovative solution. We have decided to use the DVD/CD trays from the discarded desktops, because they are easy to open and retract on command (Figure 4, Right).

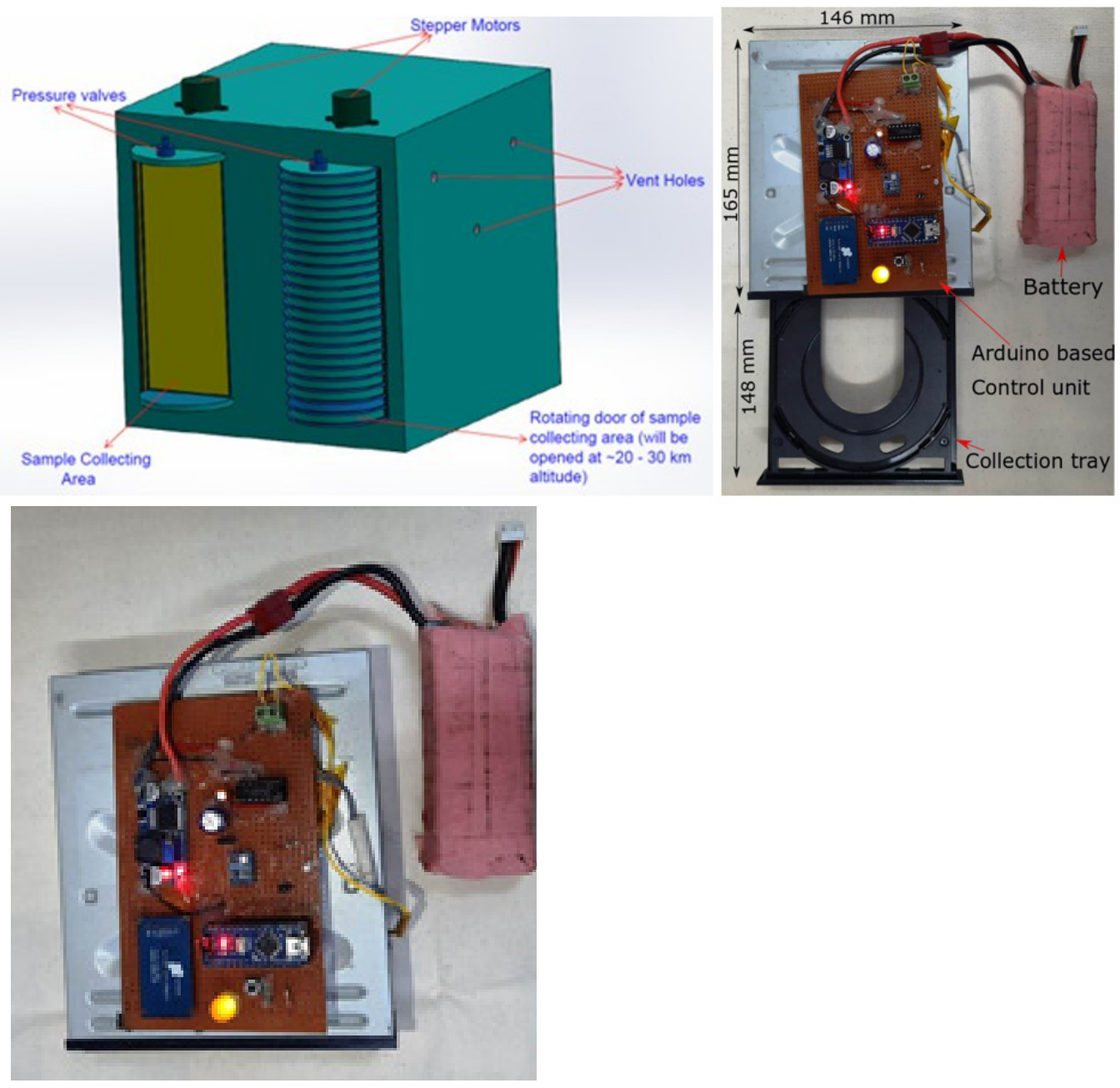

**Figure 4** *Left*: Initial mechanical design with PUF-coated plates. *Middle*: SAMPLE collector from the DVD drive with tray opened. Dimensions are shown. *Right:* SAMPLE collector with tray retracted.

## Mechanical design

### DVD drive

The SAMPLE test prototype was designed based on a 5.25-inch DVD optical drive. Mechanical design of the optical drive's gear systems and the collection tray were unaltered. The dimensions are shown in Figure 4(Middle). The casing of the SAMPLE system is white anodized Aluminum, and the interior structures are made up of polycarbonate. A lead screw is being driven by the servo motor mechanism and that leads to the in and out motion of the collection tray. The motion is constrained by a bidirectional sliding latch and lock mechanism. Opening and closing of the tray is controlled by Arduino with the altitude data inputs received from the GPS and the pressure sensor. Figure 4(Middle) shows the opened, and Figure 4(Right) shows the closed state of the SAMPLE collector. When the collection is completed, the tray will be closed and retained with the latch system so that it will not open during the decent or on the impact of landing.

In Table 2 we listed the basic specifications of the payload. Some of the SAMPLE subsystems will include:

I. Power supply: Li ion battery 2000 mAh

II. Controller: Arduino

III. Environment sensors: temperature (inside/outside), pressure, altitude

IV. GPS/Trackers

V. Radio Transmitter

VI. Aluminum alloy frame

**Table 2** SAMPLE instrument basic specifications

| **Instrument specifications** | |
|---|---|
| Operating voltage | 12 V |
| Operating temperature range | −40 to 60 ◦C |
| Mass | 300 grams |
| Collecting area | 60 cm2 |
| Collecting medium | Agar agar |
| Collecting altitude | 25 −37 km |
| Expected floating time | 60 min |
| Collecting method | Inertial impacting |

### Temperature test

Because the trays are made of polycarbonate, we have tested one in the cold chamber of the IISc Dept. of Microbiology, to simulate the conditions at high altitudes. The tray attached to the controller was kept inside the chamber at an ambient temperature of —80″ C for 2 hrs, and the command was sent 4 times to open and close the tray. According to the recorded data, the system has survived 2 hrs storage, and 4 times open-close cycling at that temperature.

### Final mechanical design – 3D print

We finally have decided that the CD tray and tray housing are to be 3D printed. The final design has electromagnetic latch, and the belt drive has been replaced with gear system made of nylon. In order to facilitate the assembly of the model, it is segmented into multiple parts:

1) Electromagnetic Latch
2) Housing Back Plate
3) Housing Base Plate
4) Main Housing
5) Tray
6) Rack – 0.6-mm module and 1.885-mm pitch
7) Rack gear – 15 teeth and 0.6-mm module
8) Motor gear – 15 teeth and 0.3-mm module
9) Transmission gear – 90 teeth and 0.3-mm module
10) BLDC motor (3V to 6V)

Figure 5 shows the updated mechanical design of the SAMPLE collector, developed to enable linear translation of a sample tray within a constrained enclosure. The design of the model was developed keeping in mind the dimensions of the outer Aluminum casing of a standard CD ROM drive. The tray motion is actuated by a rack-and-pinion mechanism, where a pinion gear mounted on a motor shaft drives a linear rack fixed beneath the tray (Figure 5(c)). This configuration provides a compact and direct means of generating translational motion from rotational input. To mitigate unintended motion due to external disturbance during balloon flight or handling during transport, a passive latch mechanism is incorporated (Figure 5(b)) along with an active latch using a solenoid. A mechanical stop is included at the end of the travel path to prevent over-extension of the tray beyond its structural limits (Figure 5(d)). These features are intended to reduce the likelihood of dislodgement or misalignment during operation.

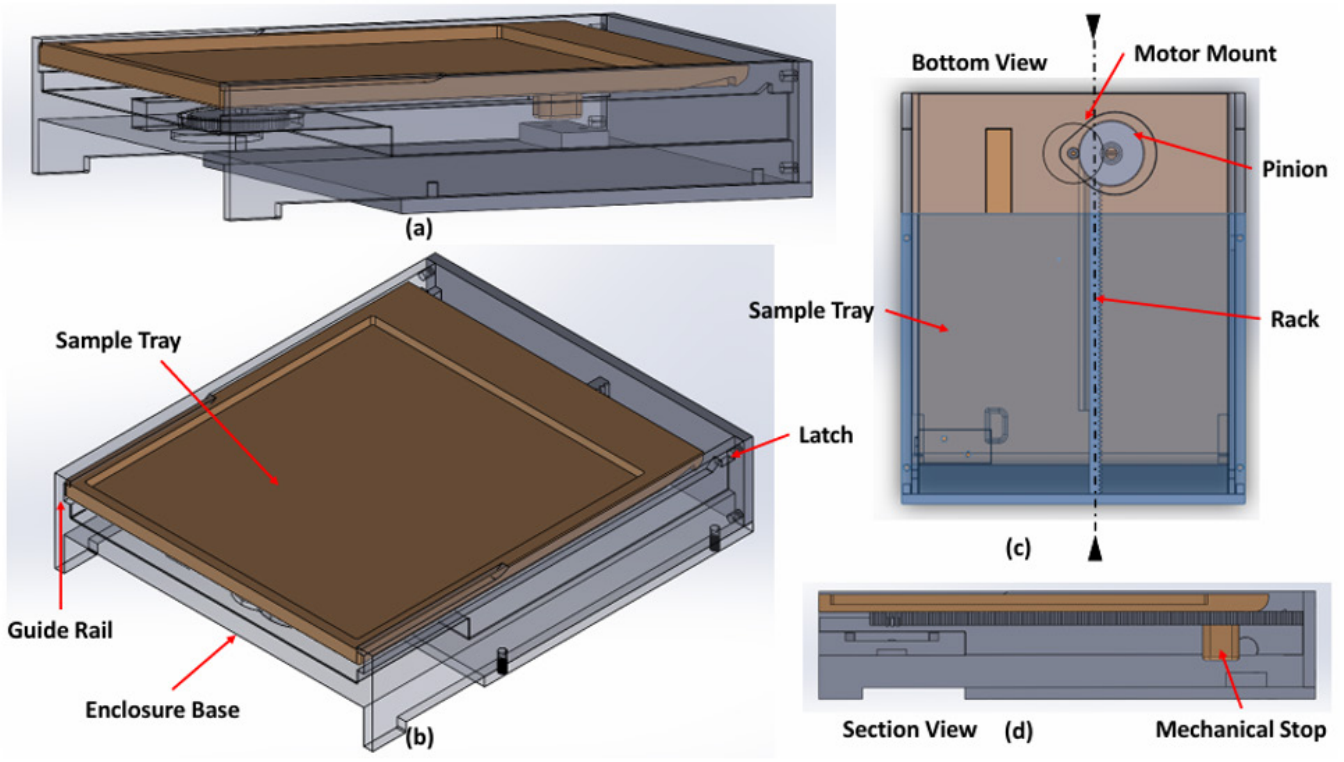

**Figure 5** Current mechanical design of SAMPLE. (a) Isometric view of the complete mechanical assembly, illustrating the spatial configuration of the internal components. (b) Isometric view, emphasizing the key structural elements: the sample tray, guide rails, and enclosure base. (c) Bottom view, showing the rack-and-pinion actuation mechanism, with the motor mount, pinion gear, and linear rack. (d) Sectional side view, depicting the integrated mechanical stop designed to constrain the maximum extension of the sample tray.

Figure 5(a) presents an isometric view of the full assembly, showing the relative positioning of internal components. Figure 5(b) illustrates the tray, guide rails, and enclosure base that together constrain motion to a single axis and support alignment. The SAMPLE collector will be constructed using materials selected for their compatibility with microbiological sampling in a balloon-borne environment, with emphasis on contamination control, sterilizability, and mechanical reliability under variable pressure and temperature conditions. The sample tray will be fabricated from polyether ether ketone (PEEK) due to its excellent chemical resistance, thermal stability, and compatibility with sterilization methods such as autoclaving, UV exposure, or plasma treatment. PEEK's low outgassing characteristics and inert surface make it suitable for microbiological applications where contamination control is critical. The tray housing will initially be prototyped using PLA for geometric verification and will be replaced in the final version with anodized Aluminum. PLA is unsuitable for microbiological use due to its porosity, hygroscopicity, and inability to withstand standard sterilization protocols. Anodized Aluminum offers better sealing performance, is chemically stable, and can be effectively sterilized using alcohol or low-temperature plasma methods. The outer casing will be machined from 6061-T6 Aluminum alloy, providing a structurally rigid and low-outgassing enclosure for flight. All interfaces are designed with elastomeric O-ring seals (e.g., fluorosilicone or silicone) to minimize air exchange and reduce the risk of pre-deployment microbial contamination. The rack-and-pinion gear system will initially be tested with nylon, and later replaced with a PEEK-based gear pair in the final configuration. Nylon is unsuitable due to its moisture absorption and limited sterilizability. PEEK offers superior mechanical stability, wear resistance, and compatibility with clean assembly environments. The microbial collection medium consists of agar gel, loaded into the sealed tray prior to flight. To prevent environmental contamination during ascent, the tray remains enclosed by a mechanically latched cover that opens only at the target altitude. The sealing and deployment mechanism is designed to minimize pre-exposure and support post-recovery microbial integrity.

## Electronics design

We have used open-source hardware board Arduino Nano as the main controller for our sample collector. The Arduino Nano uses the ATmega328 AVR micro-controller, and it has 22 general-purpose input/output pins for interfacing and control of external devices. The collector uses the BMP180 pressure sensor and a GPS module to determine the altitude; both sensors are interfaced with Arduino. The DC motor, which opens and closes the tray, is controlled by the Arduino with the help of an L293D motor controller IC. The entire system is powered by a 7V Li-ion battery; 5V is generated by using the common LM7805 linear voltage regulator, and the 12V for the motor is generated from the 7V using is an XL6009 boost converter board (Figure 6). The Arduino is programmed to open the tray when it is within the preset attitude range, and close when it is outside. A mechanical actuator switch helps the Arduino to verify if the tray is closed or open. All the events of tray opening or closing are logged on to an SD card along with the timestamps.

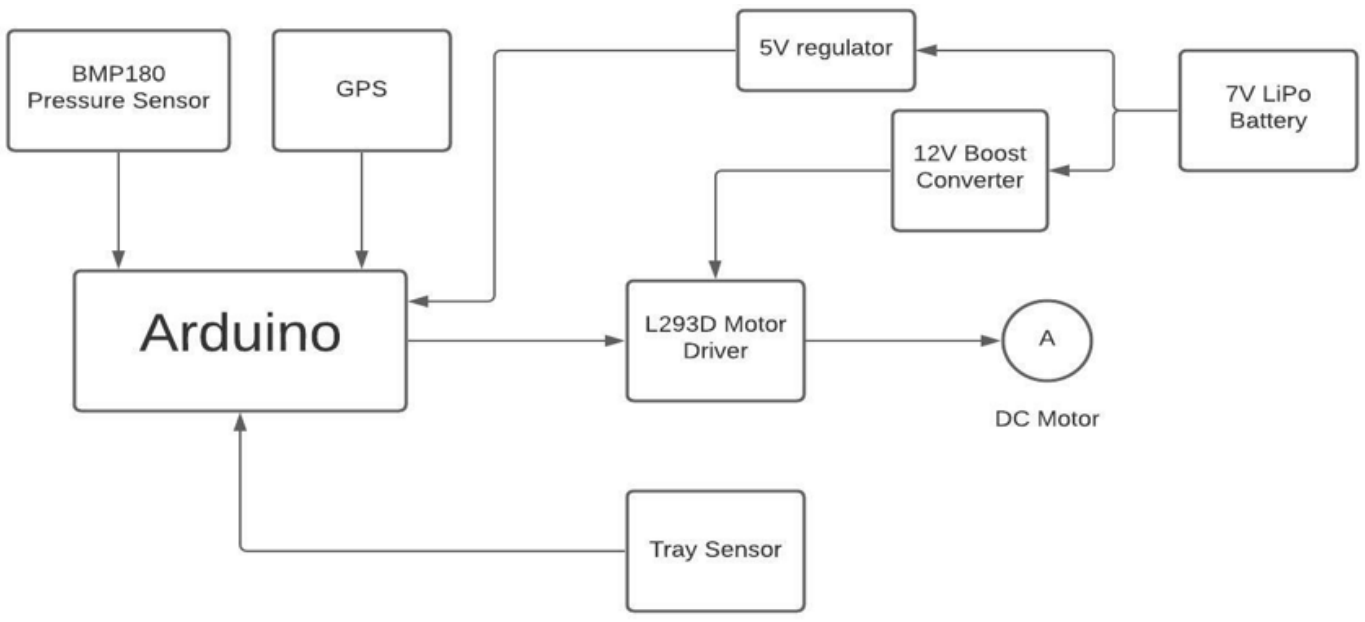

**Figure 6** Electronics block diagram of SAMPLE.

# Discussion

## Operation procedures

Before coating the collecting trays with the medium, we would need to clean them – the necessity is for not introducing even the dead/inactivated biota (sterilized), since it may interfere with later identifications. For that purpose, the collecting trays will be placed into the sonic bath filled with isopropyl alcohol (IP) solution at the M. G. K. Menon Laboratory for Space Sciences (MGKML), CREST campus, IIA. Then the trays will be placed into the sterile container/bag to deliver to the Dept. of Microbiology, IISc, for sterile coating with agar. After that, they will be kept at the MGKML, IIA, till the integration with the payload. An hour before the integration, all the surfaces will be wiped with 91% isopropyl alcohol solution. For contamination control, we will have two identically prepared extra trays, one of which will fly but not open, and one will remain closed on the ground at the launch site.

The flight will be conducted according to the usual established procedures.[31] There is an essential set of equipment that has to be present at every flight, which includes 2 sets of FTU (flight termination units), radio transmitter, GPS tracker, pressure sensors, attitude sensors, inside and outside temperature sensors, and cameras, one looking up at the balloons and one looking at the horizon. The additional payloads may include the camera to monitor the collector tray, and the Geiger-Muller counter. We plan to program the SBC to open the trays above 25 km; it was reported that at altitudes of 17–20 km, the collected CPAs (chondritic porous aggregates) contained a lot of contaminants from volcano eruptions.[72]Accordingly, the trays will be retracted once the payload crossed that altitude during the descent.

Upon the recovery of the payload, we plan to deliver it to the class 10,000 cleanroom in the MGKML, IIA, where the collecting trays will be removed from the payload box. After the outside surfaces are wiped with 91% IP solution, they will be placed into the sterile container/bag to deliver to the IISc lab.

### Investigation methods and analysis

The primary goal of the SAMPLE mission is to collect and to correctly identify the collected material. The investigation methods that we plan to employ are divided into physical and biological. The physical methods may include scanning electron microscope (SEM), mass spectrometry, and isotopic measurements of hydrogen, oxygen, nitrogen and carbon.

### Physical methods SEM

The images from the SEM will give the morphological characteristics of the collected material. There is, however, a danger that visual appearance on the images from SEM may look like 'biological entities', while there is a caveat. Inorganic processes in space are capable of producing all kinds of morphological features, even morphologies that mimic primitive living organisms. It was concluded that morphology alone cannot be used unambiguously as a tool for detection of primitive life.[73] Therefore, we consider premature the conclusions of ET biological origin of particles found by Wainwright et al.,[17] (see Figure 2 for comparisons).

### Isotopic content

Because all life on Earth preferentially uses lighter isotopes of carbon and nitrogen, we can use this fact to distinguish terrestrial organic materials the from extraterrestrial. Usually, the D/H ratio is high in space material (e.g. water) compared to the Earth ratio. If we detect different D/H ratio in our samples, or if Ni/Fe«18 (metallic meteorites), that will point to their ET origin. The high ratio of C12 to C13 would suggest a biological origin of this carbon, which is likely the result of the life forms that photosynthesize, using the Sun's energy to turn carbon dioxide into sugars. Even an elemental analysis may point to the origin: dust from Earth would very rarely have iridium or osmium in its composition.

### Biological methods

In Wainwright et al.,[17] the samples were imaged by SEM and, though believed to be some sort of ET 'biological entities', could not have been cultured. However, many terrestrial microorganisms cannot be cultured, or show only limited culturability if the environment is not suitable for them, and thus they cannot be confidently identified even being terrestrial in origin. In addition, sporification also limits the possibility of culturing, thus cultivation techniques cannot be used to verify the viability of suspected cells. If any ET biological organism is present in the samples, it is anyway not possible to detect its presence through culturing since we would not be able to determine its nutritional needs. To firmly identify whether the samples contain any living biota, we will use the modern biological techniques. If we detect any biological-looking material by the SEM investigation, the biological in- vestigation methods may include metagenomics and investigation by the Nanolive. Metagenomics is the genomics technique without the need for isolation and lab cultivation of individual species[74] and relies on the extensive existing DNA libraries.

However, even metagenomics may not be effective (or possible), as the putative ET organism may not be DNA/RNA-based, or the sequence of a terrestrial one may not be available in the library. In this regard, Nanolive is the 3D live tomographic microscopy of the cell,[75] developed in Switzerland[3]. Nanolive allows detection of cell viability without staining or culturing.

[3]http://nanolive.ch/3d-cell-explorer/

The Nanolive Technology detects low-intensity laser light propagation through the cell by means of measuring the refractive index distribution. The result is a quantitative 3D cell tomography *in vitro*, without any invasion or sample preparation, with a resolution of down to 75–90 nm.

## Conclusion

We have described the science motivation behind the experiment to collect the samples of stratospheric dust, and design and development of the payload. We are building on the experience of our group in launching the high-altitude balloons, as well as designing and manufacturing small space payloads, e.g. Chandra et al. (2024a,b).[76,77] We plan to fly the first prototype – one tray, as soon as weather and airport permissions allow. The final design will include several trays arranged in a stack, and the plan is to open them in succession at different altitudes as payload goes up and crosses various atmospheric layers to sample their dust and possible biotic content. We are currently in the process of establishing the analysis and investigation protocols, both physical and biological, as we described in this paper.

## Acknowledgments

The authors acknowledge Harshit Raj, Dhruv Gajjar, Diksha Arora, and Jagannath Prasad Sahoo from Ramaiah Institute of Technology, Bangalore, for their contributions to the current mechanical design development of the SAMPLE collector. We would like to acknowledge the help and support provided by Prof. Raghavendra Prasad, Vishnu T., S. Pawan Kumar, Nataraj, and Dr. Venkata Narra Suresh from M.G.K Menon laboratory, CREST campus, IIA.

### Author Contributions

MS wrote the main draft of the manuscript, JM is the PI of the overall project, AD and DC provided the description of the biological investigation methods. BCP contributed to the development of the DVD-based sample collector prototype and the electronics design. BGN contributed to the mechanical design development and the formulation of operational procedures. All authors contributed to the writing and editing of the manuscript. All authors have read and agreed to the published version of the manuscript.

## Conflict of Interest

The authors declare no conflict of interest. The funders had no role in the design of the study; in the collection, analyses, or interpretation of data; in the writing of the manuscript; or in the decision to publish the results.

## Funding

This research was made possible through financial support from the Indian Institute of Astrophysics (IIA) under the Department of Science and Technology (DST), India.